\documentclass[onecolumn,prb,aps,notitlepage]{revtex4-1}
\usepackage{tikz}
\usepackage{amssymb}
\usepackage{psfrag}
\usepackage{textcomp}
\usepackage{pdfsync}
\usepackage{amsmath}
\usepackage{amsfonts}
\usepackage{graphicx,bm,epstopdf}
\usepackage{subfigure}
\usepackage{comment}
\usepackage{verbatim}
\usepackage{color}
\usepackage{upgreek}

\begin{document}


\title{Mode coupling in spin torque oscillators}
\author{Olle Heinonen$^{1,2}$}
\author{Yan Zhou$^{3,4}$}
\author{Dong Li$^5$}
\affiliation{$^1$Physical Sciences and Engineering, Argonne National Laboratory, Lemont, Illinois 60439, USA\\
$^2$Department of Physics and Astronomy, Northwestern University, 2145 Sheridan Road, Evanston, Illinois 60208, USA\\
$^3$Department of Physics, The University of Hong Kong, Hong Kong, China \\
$^4$Center of Theoretical and Computational Physics, Univ. of Hong Kong, Hong Kong, China\\
$^5$Department of Physics, Centre for Nonlinear Studies, and
Beijing-Hong Kong-Singapore Joint Centre for Nonlinear and Complex
Systems (Hong Kong), Hong Kong Baptist University, Kowloon Tong,
Hong Kong, China }

\date{\today}

\begin{abstract}
A number of recent experimental works have shown that the dynamics of a single spin torque oscillator can exhibit complex behavior that stems from 
interactions between two or more modes of the oscillator. Examples are observed mode-hopping or mode coexistence. There has been some intial work indicating how the theory for a single-mode (macro-spin) spin torque oscillator should be generalized to include several modes and the
interactions between them. In the present work, we derive such a theory starting with the Landau-Lifshitz-Gilbert equation for magnetization
dynamics. We compare our result with the single-mode theory, and show how the coupled-mode theory arises as a natural extension of the
single-mode theory by including mode interactions.
\end{abstract}
\pacs{85.75.-d, 76.50.+g, 72.25.-b}
\maketitle
\section{Introduction}
Since the prediction of spin transfer torque (STT) in 1996
\cite{slonczewski1996jmmm,berger1996prb,ralph2008jmmm}, 
whereby a spin-polarized dc current exerts a torque on the local magnetization order parameter, 
there have been a wealth of theoretical and experimental investigations of phenomena driven by STT. One particular manifestation of STT is the spin torque oscillator (STO). The STO is typically realized in MgO magnetic tunnel 
junctions\cite{nazarov2008jap,deac2008np,houssameddine2009prl,zheng2010prb,muduli2012prb}, or metallic nanocontacts\cite{mancoff2006apl,rippard2004prl,silva2008jmmm}; in both of these, a dc current is driven perpendicularly to two thin stacked magnetic layers, in one of which the magnetization is relatively free to rotate, while in the other the magnetization is held fixed.  With the relative magnetization directions and current direction arranged appropriately, STT pumps energy into the STO, and by adjusting the current magntitude, this pumping can be made to cancel the intrinsic dissipative processes in the system. This gives rise to almost undamped oscillations with a very small linewidth. As STOs are potentially useful in technological applications, such as frequency generators or modulators, it is both of practical as well as of fundamental interest to understand the physics of the STO auto-oscillations. Slavin and co-workers\cite{slavin2008ieeem,tiberkevich2008prb,kim2008prl1,slavin2009ieeem} put forth a comprehensive theory valid for single-mode STOs, that is, STOs for which one mode is relevant and is excited (this is
when a macro-spin model is readily applicable). Some striking features of this theory are the effects induced by the inherent nonlinearity of the STOs, for example the behavior of the oscillator linewidth below and 
above threshold\cite{tiberkevich2008prb,kim2008prl1,slavin2009ieeem}, which is the current at which STT pumping first cancels damping and auto-oscillations are achieved. Recently, there have been several experiments demonstrating the effects of multi-mode STOs, for example mode co-existence and 
mode-hopping\cite{berkov2007prb,bonetti2012prb,bonetti2010prl,krivorotov2008prb,lee2004natmat,sankey2005prb,muduli2012prl,dumas2013prl}. Clearly, the interactions between several oscillator modes cannot be described by the single-mode theory but requires a theory that describes the interactions between collective modes, and how the behavior of the collective modes
is modified as a consequence of those interactions.
A multi-mode theory was first outlined by 
Muduli, Heinonen, and {\AA}kerman\cite{muduli2012prl,muduli2012prb,heinonen2013ieee}. In particular, these authors argued that the equations describing two-coupled modes could be mapped onto a driven dynamical system, for example used to describe semiconductor ring lasers\cite{vanderSande,beri2008prl}. It is known that in the presence of thermal noise, those equations exhibit mode-hopping in certain regions of parameter space\cite{beri2008prl}. A key observation here was that for mode-hopping to be present in a two-mode system, the time derivative of the slowly varying amplitude of one mode must be coupled linearly to the amplitude of the other mode (a so-called "back-scattering" term). Also, the authors gave some general argument for why mode-hopping is a minimum when the free layer magnetization is anti-parallel to that of the fixed layer, and then increases as the orientation moves away from anti-parallel\cite{muduli2012prl}. The purpose of the present work is to derive the equations for coupled modes from first principles (the micromagnetic Landau-Lifshitz-Gilbert equation), and to analyze the ensuing behavior of the system. We will also to compare
our results with the single-mode theory and to discuss in some detail how the present work is a generalization of the 
single-mode theory\cite{tiberkevich2008prb,kim2008prl1,slavin2009ieeem}.
We will show how the linear backscattering term arises naturally in a system with a small number, {\em e.g.,} two, of dominant modes but in which there is a bath of many modes. This bath provides effective interactions between the dominant modes when the bath is integrated out and the equations projected onto the subspace of dominant modes. 
We will also show that there are additional terms that arise when the free layer and fixed layer magnetizations are at some angle $\beta$ away from parallel or anti-parallel. These terms open up new scattering channels between modes, and therefore lead to more mode-mode interactions, and 
provide a mechanism for the observed\cite{muduli2012prl} increased mode-hopping away from parallel or anti-parallel free and fixed layer magnetizations. The geometry we use is specifically adapted for a magnetic tunnel junction
with in-plane magnetization and an in-plane external field, although we will also discuss some other geometries. 

\section{Micromagnetic equations}
\begin{figure}
\includegraphics*[width=.45\textwidth]{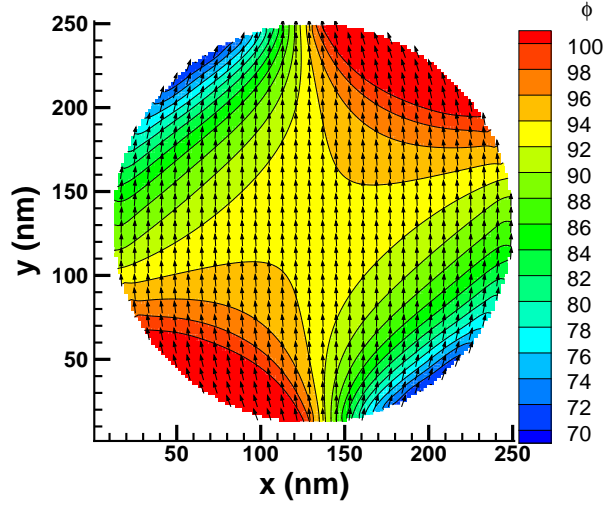}
\caption{(Color online) Micromagnetic equilibrium magnetization in the free layer of a circular magnetic tunnel junction of diameter 240 nm. The pinned and
reference layer magnetizations (not shown) are at $45^\circ$ and $-135^\circ$ degrees to the $x$-axis, and there is
and external field of 450~Oe applied at $85^\circ$ to the $x$-axis. The color scale shows the angle $\phi$ between the local
magnetization and the $x$-axis.}
\label{fig:MTJ_FL}
\end{figure}

Our starting point is a soft ferromagnetic system, for example a thin film. We describe the local magnetization by a director $\hat m_i$ for discrete
sites $i=1,2,\ldots N$, with $|\hat m_i|=1$. The LLG equation including damping and spin torque is then
\begin{equation}\label{eqn:LLG_1}
\frac{d\hat m_i}{dt} = -\gamma \hat m_i\times {\mathbf H}_{{\rm eff},i}-\frac{\gamma\alpha}{1+\alpha^2}\hat m_i\times 
\left[\hat m_i\times {\mathbf H}_{{\rm eff},i}\right]+\gamma a_J\hat m_i\times\left[\hat m_i\times \hat M\right].
\end{equation}
Here, $\gamma$ is the gyromagnetic ratio, $\alpha\ll 1$ the dimensionless damping, $a_J$ the effective field due to STT, and $\hat M$ the (uniform) magnetization direction of the fixed layer; the effective field ${\mathbf H}_{{\rm eff},i}$ includes exchange, demagnetizing fields, and an external applied field ${\mathbf H}_{\rm ext}=H_{\rm ext}\hat x$. We will not here include Oersted fields generated by the currents in the system as they are not important for the present analysis, 
although it has shown that these fields play an important role in the interactions between certain modes in
nano-contact STOs\cite{dumas2013prl}. We are also ignoring the so-called field-like, or perpendicular, spin torque\cite{zhang2002prl} as this can be absorbed
into the definition of the external field. We shall combine exchange and demagnetizing fields into a single field
${\mathbf H}_{{d},i}$ and note that in general we can write
\begin{equation}
H_{d,i,\delta}=\sum_{i',\epsilon}D_{i,i';\delta,\epsilon}m_{i',\epsilon},\,\,\delta,\epsilon=x,y,z,\label{eqn:demag}
\end{equation}
where $D_{i,i';\delta,\epsilon}$ is a generalized demagnetizing tensor that includes near-neighbor exchange.
We shall also assume that the anisotropy is negligible, and that the equilibrium free layer magnetization direction is aligned with the external field along the $\hat x$ axis. This applies, for example, to magnetic tunnel junctions with an in-plane external field, or to systems with a large
($H_{\rm ext}\gg4\pi M_S$)
external
field perpendicular to the planes of the magnetic layers. Figure \ref{fig:MTJ_FL} depicts the equilibrium magnetization in the free layer of a circular
magnetic tunnel junction STO of diameter $d=240$~nm obtained from micromagnetic simulations with parameters appropriate for the systems in Ref.~\onlinecite{muduli2011prb}. In the figure, the pinned layer and reference layer magnetizations are approximately
(these layers are also treated micromagnetically) at $45^\circ$ and $-135^\circ$ degrees to the $x$-axis, and there is 
an external field of magnitude 450~Oe applied in the $xy$-plane at $85^\circ$ to the $x$-axis (in an actual magnetic tunnel juction, there
are three magnetic layers, and the reference layer is the one next to the free layer and is responsible for the spin transfer torque). 
The figure shows that while the magnetization is not perfectly aligned with the external field
everywhere, the maximum deviation is small, about $15^\circ$, and an expansion in deviations from alignment with the
uniform external field is reasonable and should converge rapidly. It is of course easy to generalize to an arbitrary local 
effective field direction and with a non-uniform 
equilibrium magnetization in the direction of the local net magnetic field, but that does introduce more variables. We prefer to keep the discussion relatively simple and transparent, and also with the specific application in mind of a magnetic tunnel junction STO with an in-plane
external field, or a nanocontact STO with a strong out-of-plane external field.
For simplicity, we will use units in which $\gamma=1$, and because $\alpha\ll1$ we will ignore terms of order $\alpha^2$. In terms of components, the LLG equation is then
\begin{eqnarray}
\frac{dm_{x,i}}{dt} & = & -\left(m_{i,y}H_{d,i,z}-m_{i,z}H_{d,i,y}\right)+\alpha\left(1-m_{i,x}^2\right)\left(H_{\rm ext}+H_{d,i,x}\right)
-\alpha m_{i,x}\left( m_{i,y}H_{d,i,y}+m_{i,z}H_{d,i,z}\right)\nonumber\\
&&-a_J\left(1-m_{i,x}^2\right)\cos\beta-a_J m_{i,x}m_{i,y}\sin\beta\label{eqn:LLG_comp_x}\\
\frac{dm_{y,i}}{dt} & = & -\left[m_{i,z}\left(H_{d,i,x}+H_{\rm ext}\right)-m_{i,x}H_{d,i,z}\right]+\alpha\left(1-m_{i,y}^2\right)H_{d,i,y}
-\alpha m_{i,y}\left[ m_{i,x}\left(H_{d,i,x}+H_{\rm ext}\right)+m_{i,z}H_{d,i,z}\right]\nonumber\\
&&+a_J\left(1-m_{i,y}^2\right)\sin\beta-a_J m_{i,x}m_{i,y}\sin\beta\label{eqn:LLG_comp_y}\\
\frac{dm_{z,i}}{dt} & = & -\left[m_{i,x}H_{d,i,y}-m_{i,y}\left(H_{\rm ext}+H_{d,i,x}\right)\right]+\alpha\left(1-m_{i,z}^2\right)H_{d,i,z}
-\alpha m_{i,z}\left[ m_{i,x}\left(H_{d,i,x}+H_{\rm ext}\right)+m_{i,y}H_{d,i,y}\right]\nonumber\\
&&+a_J m_{i,z}\left(m_{i,x}\cos\beta-m_{i,y}\sin\beta\right).\label{eqn:LLG_comp_z}
\end{eqnarray}

Next, we introduce the generalized local non-linear Holstein-Primakoff transformation of 
Slavin and co-workers\cite{slavin2008ieeem,slavin2009ieeem} that transforms the two degrees of freedom of each director $\hat m_i$ to a complex variable $c_i$,
\begin{equation}
c_i=\frac{m_{y,i}-i m_{z,i}} {\sqrt{2\left(1+m_{x,i}\right)}},
\label{eqn:HP_1}
\end{equation}
with the inverse transformation
\begin{equation}
\hat m_i=\sqrt{1-|c_i|^2} \left[ \left(c_i+c^*_i\right)\hat y+i\left(c_i-c_i^*\right)\hat z\right]+\left(1-2|c_i|^2\right)\hat x.
\label{eqn:HP_inverse}
\end{equation}
In terms of the local variables $c_i$, we can write the fields ${\mathbf H}_{d,i}$ as
\begin{eqnarray}
H_{d,i,x} & = & \sum_{i'}\left[D_{i,i';x,x}\left(1-2|c_{i'}|^2\right)+D_{i,i';x,y}\sqrt{1-|c_{i'}|^2}\left(c_{i'}+c_{i'}^*\right)
+D_{i,i';x,z}i\sqrt{1-|c_{i'}|^2}\left(c_{i'}-c_{i'}^*\right)\right]\label{eqn:demag_x}\\
H_{d,i,y} & = & \sum_{i'}\left[D_{i,i';y,x}\left(1-2|c_{i'}|^2\right)+D_{i,i';y,y}\sqrt{1-|c_{i'}|^2}\left(c_{i'}+c_{i'}^*\right)
+D_{i,i';y,z}i\sqrt{1-|c_{i'}|^2}\left(c_{i'}-c_{i'}^*\right)\right]\label{eqn:demag_y}\\
H_{d,i,z} & = & \sum_{i'}\left[D_{i,i';z,x}\left(1-2|c_{i'}|^2\right)+D_{i,i';z,y}\sqrt{1-|c_{i'}|^2}\left(c_{i'}+c_{i'}^*\right)
+D_{i,i';z,z}i\sqrt{1-|c_{i'}|^2}\left(c_{i'}-c_{i'}^*\right)\right]\label{eqn:demag_z}
\end{eqnarray}

By multiplying Eq.~(\ref{eqn:LLG_comp_z}) by $i$ and subtracting the result from Eq.~(\ref{eqn:LLG_comp_y}), and also using 
$m_{x,i}=1-2|c_i|^2$ from Eq.~(\ref{eqn:HP_inverse}) in Eq.~(\ref{eqn:LLG_comp_x}),
we obtain the following equations:
\begin{eqnarray}
\frac{d}{dt}\left[2\sqrt{1-|c_i|^2}c_i\right] & = & \left(1-2|c_i|^2\right)\left(H_{d,i,z}+iH_{d,i,y}\right)-
2i\sqrt{1-|c_i|^2}c_i\left(H_{\rm ext}+H_{d,i,x}\right)\nonumber\\
& & -2\alpha \sqrt{1-|c_i|^2} c_i \left[ m_{i,x}\left( H_{\rm ext}+H_{d,i,x}\right)+m_{i,y}H_{d,i,y}+m_{i,z}H_{d,i,z}\right]
\nonumber\\
&&+\alpha\left(H_{d,i,y}-iH_{d,i,z}\right)+2a_J\sqrt{1-|c_i|^2}c_i\left(m_{i,x}\cos\beta-m_{i,y}\sin\beta\right)+a_J\cos\beta\label{eqn:dc_dt_1}\\
-2\frac{d}{dt}|c_i|^2 & = & -\sqrt{1-|c_i|^2}\left[\left(c_i+c_i^*\right)H_{d,i,z}-i\left(c_i-c_i^*\right)H_{d,i,y}\right]
+4\alpha\left(1-|c_i|^2\right)|c_i|^2\left(H_{\rm ext}+H_{d,i,x}\right)\nonumber\\
&&-\alpha\left(1-2|c_i|^2\right) \sqrt{1-|c_i|^2}\left[\left(c_i+c_i^*\right)H_{d,i,y}+i\left(c_i-c_i^*\right)H_{d,i,z}\right]\nonumber\\
& & -4a_J\cos\beta\left(1-|c_i|^2\right)|c_i|^2-a_J\left(1-2|c_i|^2\right)\sqrt{1-|c_i|^2}\left(c_i+c_i^*\right)\sin\beta.\label{eqn:dc_dt_2}
\end{eqnarray}
Finally, we combine Eqs.~(\ref{eqn:dc_dt_1} - \ref{eqn:dc_dt_2}) and use
\begin{equation}
\frac{d}{dt}\left[2\sqrt{1-|c_i|^2}c_i\right]=2\sqrt{1-|c_i|^2}\frac{dc_i}{dt}-\frac{c_i}{\sqrt{1-|c_i|^2}}\frac{d|c_i|^2}{dt}
\end{equation} 
to obtain
\begin{eqnarray}
\frac{dc_i} {dt}&=&-\frac{(i H_{d,i,y}+H_{d,i,z}) (-2+3 |c_i|^2)}{4 \sqrt{1-|c_i|^2}}-i (H_{\rm ext}+H_{d,i,x}) c_i-\frac{(i H_{d,i,y}-H_{d,i,z}) c_i^2}{4 \sqrt{1-|c_i|^2}}\nonumber\\
&&+\alpha\frac{(H_{d,i,y}-i H_{d,i,z}) \left[{2+|c_i|^2 (-3+2 |c_i|^2)}\right] }{4 \sqrt{1-|c_i|^2}}
-\alpha c_i(H_{\rm ext}+H_{d,i,x}) (1-|c_i|^2)+\alpha c_i^2\frac{(H_{d,i,y}+i H_{d,i,z}) (-3+2 |c_i|^2) }{4 \sqrt{1-|c_i|^2}}\nonumber\\
&&+a_J\frac{\left[{2+|c_i|^2 (-3+2 |c_i|^2)}\right] \sin[\beta ]}{4 \sqrt{1-|c_i|^2}}+a_J c_i (1-|c_i|^2) \cos[\beta ]+a_J c_i^2\frac{(-3+2 |c_i|^2) \sin[\beta ]}{4 \sqrt{1-|c_i|^2}}\label{eqn:dc_dt_3}
\end{eqnarray}
The somewhat unpleasant-looking Eq.~(\ref{eqn:dc_dt_3}), together with Eqs.~(\ref{eqn:demag_x} - \ref{eqn:demag_z}), is equivalent to the traditional LLG equation and describes the full non-linear magnetization motion in the presence of damping and (in-plane) STT. The advantage of this form compared to the LLG form is that it allows for a systematic expansion in powers of $c_i$ to derive the effective time-evolution of coupled modes. For STT auto-oscillators, we must also include a non-linear dependence of the damping $\alpha$ on the oscillator power\cite{slavin2008ieeem} -- otherwise we will not get stable oscillations above threshold -- and we will in general write $\alpha=\alpha_G(1+q_1\xi^2)$, where $\xi$ is a dimensionless measure of the oscillator energy.

\section{Non-conservative torques}
We will first use Eq.~(\ref{eqn:dc_dt_3}) to analyze the effect of the non-conservative torques on an auto-oscillator. By assumption, the magnetization motion is oscillatory with a period $\tau$, and the magnetization amplitude remains constant or invariant under long times. The dissipation of the system is described by the time-rate of change of the oscillator power, proportional to $\sum_i|c_i|^2$. For simplicity, and ease of notation, we now use a macro-spin model\cite{li2003prb} with a single amplitude $c$. The rate of dissipation is then given by
\begin{eqnarray}
\frac{d}{dt}|c|^2 & =& \sqrt{1-|c|^2}\left[\Re(c)H_{d,z}+\Im(c)H_{d,y}\right]\nonumber\\
&&+\frac{1}{\sqrt{1-|c|^2}} \left\{\left[\Re(c)\left(a_J\sin\beta+\alpha H_{d,y}\right)-\alpha H_{d,z}\Im(c)\right]
\left[1+|c|^2\left(-3+2|c|^2\right)\right]\right\}\nonumber\\
&&-2\left(1-|c|^2\right)|c|^2\left[\alpha\left( H_{\rm ext}+H_{d,x}\right) -a_J\cos\beta\right].\label{eqn:disspation_1}
\end{eqnarray}
Here $\Re(c)$ and $\Im(c)$ are the real and imaginary parts of $c$, respectively.
For simplicity, we also assume that the demagnetizing tensor is diagonal (this is not a very drastic assumption for elliptical systems)
so that $H_{d,y}=-Ym_y=-2Y\sqrt{1-|c|^2}\Re(c)$ and $H_{d,z}=-Zm_z=2Z\sqrt{1-|c|^2}\Im(c)$, with $Y$ and $Z$ real numbers. 
Then $\frac{1}{\sqrt{1-|c|^2}}\Re(c)H_{d,y}=-2Y\left[\Re(c)\right]^2$ and $\frac{1}{\sqrt{1-|c|^2}}\Im(c)H_{d,z}=2Z\left[\Im(c)\right]^2$.
We demand that averaged over a period $\tau$, the dissipation is zero, so that
\begin{equation}
\langle\frac{d}{dt}|c|^2\rangle=\frac{1}{\tau}\int_t^{t+\tau}\frac{d}{dt}|c(t')|^2\,dt'=0.
\end{equation}
For an oscillatory motion we can assume
\begin{equation}
\frac{1}{\tau}\int_t^{t+\tau}\left[\Re(c(t'))\right]^2\,dt'=\overline c^2 >0,
\end{equation}
which defines the number $\overline c$.
The motion will in general be eccentric with eccentricity $\epsilon$, so
\begin{equation}
\frac{1}{\tau}\int_t^{t+\tau}\left[\Im(c(t'))\right]^2\,dt'\approx\epsilon\overline c^2.
\end{equation}
Futhermore, for oscillatory motion we have
\begin{equation}
\frac{1}{\tau}\int_t^{t+\tau}\Re[c(t')]\,dt'\approx0\approx\frac{1}{\tau}\int_t^{t+\tau}\Re[c(t')]\Im[c(t')]\,dt'.
\end{equation}
To ${\cal O}(c^2)$ we then obtain
\begin{equation}
\langle\frac{d}{dt}|c|^2\rangle=-(Y+\epsilon Z)\alpha_G-\left[\alpha_G\left(H_{\rm ext}+\overline H_{d,x}\right)-a_J\cos\beta\right]\left[1+\epsilon\right]
=0,
\label{eqn:single_mode_thresh}
\end{equation}
where $\overline H_{d,x}$ is the equilibrium demagnetizing field in the $x$ direction, and a factor of $\overline c^2$ cancels out. 
This analysis allows us to draw three 
conclusions: ({\em i}) the time-dependent demagnetizing fields enhance the average dissipation by a factor of $(Y+\epsilon Z)\alpha$;  
({\em ii}) 
only the {\em average} dissipation is zero, but not the instantaneous dissipation, and during some fraction of a period net energy is pumped into the system, and during some other fraction of a period net energy is 
dissipated\cite{muduli2012prl}; ({\em iii}) as $\beta \to \pi/2$ the pumping through STT
becomes less and less effective to offset dissipative losses and it becomes in general impossible
to obtain self-sustained auto-oscillations. The threshold current is the current at which the average dissipation is zero. 
From Eq.~(\ref{eqn:single_mode_thresh}) we see that the spin torque effective field $a_J$, which is proportional to the current, and $\cos\beta$
enters as a product. This implies that the threshold current increases as $1/\cos\beta$ as the equilibrium magnetization direction is rotated
away from the direction of the reference layer. An increase in threshold current with angle $\beta$ has indeed been observed 
experimentally\cite{muduli2012prl}. The dependence on the product $a_j\cos\beta$ also implies an invariance: if a decrease in $\cos\beta$ is
offset by an increase in $a_J$ by increasing the current such that the product $a_J\cos\beta$ is constant the system is invariant. This is
{\em not} consistent with experimental observations, where, for example, mode-hopping increases dramatically as $\cos\beta$ is 
decreased\cite{muduli2012prl}. As we will argue below, this can only be caused by by the apperance of terms in $a_J\sin\beta$ in the
coupled-mode equations.

\section{Comparison with singe-mode theory}
It is instructive to compare Eq.~(\ref{eqn:dc_dt_3}) with the single-mode theory. To this end, we assume that there is a single macro-spin,
and also assume that the demagnetizing tensor is diagonal, $D_{xx}=-X$, $D_{yy}=-Y$, and $D_{zz}=-Z$. Inserting this, and the transformations
Eq.~(\ref{eqn:HP_inverse}) and expanding to third order, we obtain after a little algebra
\begin{eqnarray}
\frac{dc}{dt} & = & -\frac{i}{2}Y(c+c^*)\left(1-\frac{3}{2}|c|^2-\frac{1}{2}c^2\right)-\frac{i}{2}Z(c-c^*)\left(1-\frac{3}{2}|c|^2+\frac{1}{2}c^2\right)
-icH_{\rm ext}+icX(1-2|c|^2)\nonumber\\
&&-\frac{\alpha}{2}Y(c+c^*)\left(1-\frac{3}{2}|c|^2-\frac{3}{2}c^2\right)-\frac{\alpha}{2}Z(c-c^*)\left(1-\frac{3}{2}|c|^2+\frac{3}{2}c^2\right)
-\alpha cH_{\rm ext}(1-|c|^2)+\alpha cX(1-3|c|^2)\nonumber\\
&&+\frac{a_J}{2}\sin\beta\left(1-|c|^2-\frac{3}{2}c^2\right)+a_Jc(1-|c|^2)\cos\beta.\label{eqn:single_mode_1}\end{eqnarray}
The linearized conservative part of this equation is
\begin{eqnarray}
\frac{dc}{dt} & = &-\frac{i}{2}Y(c+c^*)-\frac{i}{2}Z(c-c^*)-iH_{\rm ext}c+iXc\nonumber\\
&=&-i\left\{
\left[
\frac{1}{2}\left(Y+Z\right)+\left(H_{\rm ext}-X\right)
\right]c+\frac{1}{2}\left(Y-Z\right)c^*
\right\},\label{eqn:linearized}\end{eqnarray}
with eigenvalue $\omega^2=\left[\left(H_{\rm ext}-X\right)+Y\right]\left[\left(H_{\rm ext}-X\right)+Z\right]$. 
We can diagonalize Eq.~(\ref{eqn:linearized}) by introducing a complex variable $a$ such that
$c=ua-v{a}^*$ 
with 
\begin{eqnarray}
u & = & \sqrt{
\frac{\left[H_{\rm ext}-X+\frac{1}{2}\left(Y+Z\right)\right]+\omega}{2(H_{\rm ext}-X)+(Y+Z)}}=\sqrt{\frac{b+\omega}{2b}}
\label{eqn:u_def}\\
v & = & \sqrt{
\frac{\left[H_{\rm ext}-X+\frac{1}{2}\left(Y+Z\right)\right]-\omega}{2(H_{\rm ext}-X)+(Y+Z)}}=\sqrt{\frac{b-\omega}{2b}},
\label{eqn:v_def}
\end{eqnarray}
where $b=(H_{\rm ext}-X)+\frac{1}{2}(Y+Z)$. We also define $n=\frac{1}{2}(Y-Z)$ so that $\omega^2=b^2-n^2$.
Inverting the relation between $a$ and $c$ we get
\begin{equation}
a=\frac{uc+vc^*}{u^2-v^2}\end{equation}
and $u^2-v^2=\frac{1}{\omega}\left[H_{\rm ext}-X+\frac{1}{2}(Y+Z)\right]=b/\omega$ and $u^2+v^2=1$.
With this in mind, we can re-write Eq.~(\ref{eqn:single_mode_1}) as
\begin{eqnarray}
\frac{dc}{dt} & = & -i\left[bc+nc^*\right]
\left[1-\frac{3}{2}|c|^2\right]\left[1-i\alpha\right]\nonumber\\
&& -\frac{i}{2}\left(3H_{\rm ext}+X\right)c|c|^2 
+\frac{i}{2}nc^3+\frac{i}{4}\left(Y+Z\right)c|c|^2
\nonumber\\
&&-\frac{\alpha}{2}\left( H_{\rm ext}+3X\right)c|c|^2
+\frac{3\alpha}{2}nc^3+\frac{3\alpha}{4}\left(Y+Z\right) c|c|^2
\nonumber\\
&&+\frac{a_J}{2}\sin\beta\left(1-|c|^2-\frac{3}{2}c^2\right)+a_J\cos\beta c(1-|c|^2).\label{eqn:single_mode_2}
\end{eqnarray}
The conservative parts of the equation of motion for $c$ can be derived from a hermitian 
Hamiltonian $H$ by $dc/(dt)=-i\partial H/(\partial c^*)$, where
\begin{equation}
H=b|c|^2+\frac{n}{2}\left(c^{*2}+c^2\right)-\frac{n}{2}\left(c^3c^*+cc^{*3}\right)+\frac{1}{4}\left[3H_{\rm ext}+X-\frac{1}{2}(Y+Z)\right]|c|^4.
\label{eqn:hamiltonian}
\end{equation}
We note that this is different in structure from the Hamiltonian of 
Slavin and Tiberkevich (Eq.~(3.16) in Ref.~\onlinecite{slavin2008ieeem}) in that Eq.~(\ref{eqn:hamiltonian}) contains no cubic terms. This just stems from our choice of geometry in which the magnetization is aligned with the external field; if
the external field has components transverse to the equilibrium magnetization, quadratic terms will appear in the equation of motion 
Eq.~(\ref{eqn:dc_dt_3}) with ensuing cubic terms in the Hamiltonian.

Finally, we introduce a variable $m=\frac{1}{2}(Y+Z)$, expand $\alpha=\alpha_G(1+q_1|a|^2)$ and write Eq.~(\ref{eqn:single_mode_2}) in terms of $a$ (see Appendix A for details):
\begin{eqnarray}
\frac{da}{dt} & = & -i\omega a -\omega a\alpha_G\left[1+q_1|a|^2\right]+\frac{3i\omega}{2}a
\left[|a|^2-\frac{n}{2b}\left( a^2+ a^{*2}\right)\right][1-i\alpha]\nonumber\\
&&-\frac{i}{2}\left[3H_{\rm ext}+X-m\right]\left[
\frac{2b^2+n^2}{2b\omega}a|a|^2-\frac{3n}{2\omega}a^*|a|^2-\frac{n}{2\omega}a^3+\frac{n^2}{2b\omega}{a}^{*3}
\right]\nonumber\\
&&+\frac{i n}{2}\left[
\frac{3n^2}{2b\omega}a^*|a|^2-\frac{3n}{2\omega}a|a|^2+\frac{b^2
+\omega^2}{2b\omega}a^3-\frac{n}{2\omega}{a}^{*3}
\right]
\nonumber\\
&&-\frac{\alpha}{2}\left[ H_{\rm ext}+3X-3m\right]\left[
a|a|^2-\frac{n}{2b}\left( a^3+a^*|a|^2\right)\right]\nonumber\\
&&+\frac{3\alpha n}{2}\left[ a^3+\frac{n}{2b}{a}^{*3}-\frac{3n}{2b}a|a|^2\right]
\nonumber\\
&&+\frac{a_J}{2}\frac{b}{\omega}\sin\beta\sqrt{\frac{b+n}{b}}\left[1+\frac{3n-2b}{2b}|a|^2+\frac{5n-6b}{4b}a^2
-\frac{n}{4b}{a}^{*2}\right]
\nonumber\\
&&+a_J\cos\beta a\left[
1-|a|^2+\frac{n}{2b}\left( a^2-a^{*2}\right)\right].
\label{eqn:single_mode_3}
\end{eqnarray}
Equation~(\ref{eqn:single_mode_3}) looks rather complicated in that it contains all cubic terms in $a$ and $a^*$, and not just
cubic terms of the form $a|a|^2$. However, for the special case $n=0$ (in which case $\omega=b$), the magnetization motion is circularly polarized
($c=a$) and the equation of motion for $a$ attains a very simple form,
\begin{eqnarray}
\frac{da}{dt} & = & -i\omega a -\omega a\alpha_G\left[1+q_1|a|^2\right]+\frac{3i\omega}{2}a
|a|^2[1-i\alpha]
-\frac{i}{2}\left[3H_{\rm ext}+X-m\right]
a|a|^2\nonumber\\
&&-\frac{\alpha}{2}\left[ H_{\rm ext}+3X-3m\right]
a|a|^2
+\frac{a_J}{4}\sin\beta\left[2|a|^2-3a^2
\right]
+a_J\cos\beta a\left[
1-|a|^2\right].
\label{eqn:single_mode_4}
\end{eqnarray}
It is clear that the general equation Eq.~(\ref{eqn:single_mode_3}) has an expansion in $|a|^2$ only if $n=0$, which means $Y=Z$ or $Y=Z=0$, {\em and} if $\sin\beta=0$. This is
the case, for example, for the magnetization in a continuous thin film and the external field ($H_{\rm ext} > 4\pi M_S$) applied perpendicularly to the film plane, in 
which case $X=4\pi M_S$, and with the fixed layer magnetization perpendicular to the plane, so that $\sin\beta=0$, or for
an in-plane circular magnetic tunnel junction with the external field aligned with the fixed layer magnetization direction. But for a general object, such as a non-circular patterned magnetic tunnel junction or a nanocontact, $Y\not=Z$, and an expansion
in $a$ will contain all terms in $a$ and $a^*$. The origin of these latter terms in the conservative torque
comes from the time-derivative of $|c|^2$, Eq.~({\ref{eqn:dc_dt_2}). The conservative third-order terms can be derived from a quartic
Hamiltonian. For a single-mode (macrospin) STO only the fourth-order term in $|a|^4$ in the Hamiltonian 
is relevant as other fourth-order terms do not
lead to energy-conserving processes, and this term generates the term in $a|a|^2$ in the equation of motion 
responsible for the non-linear frequency shift, which then depends only on the oscillator power (or energy). 
For multi-mode systems, however, the cubic and higher-order terms both in the conservative and non-conservative parts of the
equation of motion have to be considered, and they lead to energy flow between different modes of the system ({\em e.g.,} mode-hopping) as
well as modified damping.

\section{Mode equations}
We will now proceed to derive effective equations for the slow time-behavior of modes in the presence of coupling. By slow we here mean time-scales much larger than characteristic resonance periods of the system, which are of the order of 0.1 - 1 ns; we will make this more precise below. The strategy we will use is very similar to that used in 
physical optics, where Ansatz solutions are typically made for solutions that are linear combinations of eigenmodes, and the coefficients are slowly time dependent\cite{sorel2002optlet,sorel2003ieee}. To be specific, we will consider a system
where two modes, labeled 1 and 2, are dominant. This can be a system such as the one in Ref.~\onlinecite{muduli2011prb} where
the uniform FMR mode is most strongly excited by the spin transfer torque, but where there is mode hopping between
this mode and another mode, indicative of strong couplings between those modes, or it can be a system that is
close to a mode crossing, such as the system in Ref.~\onlinecite{muduli2012prb}. We shall also assume that $\omega_1$ and $\omega_2$
are the two lowest frequencies in the system, and that there is a finite gap between $\omega_i$ and $\omega_n$, where
$i=1,2$ and $n=3,4,\ldots$. Figure~\ref{fig:MTJ_spectrum} depicts an experimental example of such a spectrum.
\begin{figure}[t!]
\includegraphics*[width=.45\textwidth]{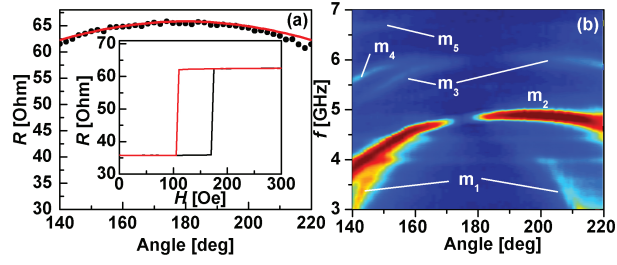}
\caption{(Color online)(a) Experiment (circles) and calculated (solid line) resistance versus in-plane field angle $\varphi$ at $H$=450~Oe, where $\varphi=180^\circ$ corresponds to anti-parallel alignment of the free layer and the reference layer.
The inset shows magnetoresistance loop measured at $\varphi=180^\circ$. (b)~Map of power (dB) vs frequency and in-plane field angle for $I$=8~mA and $H$=450~Oe showing the free layer modes. From Ref.~\onlinecite{muduli2012prl} (reproduced with permission).}\label{fig:MTJ_spectrum}
\end{figure}

In order to express a solutions in modes, we must specify a reference system for which the modes are well defined. We will use the linearized conservative system as a reference. This system is described by
\begin{equation}
\frac{dc_i}{dt}=-\frac{1}{2}(i H_{d,i,y}+H_{d,i,z})-i (H_{\rm ext}+H_{d,i,x}) c_i\label{eqn:linearized_1}.\end{equation}
This is a linear conservative system that admits a family of $N$ orthonormal (complex) eigenvectors $\left\{d_{n}\right\}$, $n=1,2\ldots,N$, that span the range of the linearized equation, with real eigenvalues $\omega_n$:
\begin{equation}
\frac{d}{dt}  d_{i,n}(t) =\frac{d}{dt} d_{i,n} {\rm e}^{-i\omega_nt}=-i\omega_n  d_{i,n} {\rm e}^{-i\omega_nt}=-\frac{1}{2}(i H_{d,i,y}+H_{d,i,z}) -i (H_{\rm ext}+H_{d,i,x}) d_{i,n}\label{eqn:linearized_2},\,i=1,2\ldots, N.\end{equation}
We will use short-hand notation $|d_n\rangle$ and $\langle d_n|$ for the vector $(d_{1,n},d_{2,n},\ldots,d_{N,n})$ and its adjoint, respectively; the 
orthogonality and completeness can then be written
\begin{equation}
\langle d_{n}| d_{n'}\rangle=\sum_i d_{i,n}^*d_{i,n'}=\delta_{n,n'},\label{eqn:orthonorm}
\end{equation}
and
\begin{equation}
\frac{1}{2\pi}\int_{-\infty}^\infty {\rm e}^{-i(\omega_n-\omega_{n'})t}\,dt=\delta_{n,n'}.\label{eqn:freq_comp}\end{equation}
We then expand the magnetization motion in the basis $\left\{ {d_n}\right\}$
\begin{equation}
|c(t)\rangle=\sum_n A_n(t) | d_n\rangle {\rm e}^{-i\omega_n t},
\label{eqn:expansion}
\end{equation}
where we assume that the complex coefficients $A_n(t)$ have a slow time dependence, that is, the time-scale of variation of 
$A_n(t)$ is much larger than $(\omega_m)^{-1}$ for all $n,m=1,2,\ldots,N$. We insert the expansion Eq.~(\ref{eqn:expansion}) in the equation for the time
evolution of the magnetization, Eq. ~(\ref{eqn:dc_dt_3}) and project with $\langle d_{n'}|$. The left-hand side of the resulting equation then
becomes simply $dA_{n'}(t)/(dt) {\rm e}^{-i\omega_{n'}t} -i\omega_{n'}A_{n'}(t) {\rm e}^{-i\omega_{n'}t}$. The second term proportional to
$\omega_{n'}$ is canceled by construction by the linear conservative part on the right-hand side of the equation. The remainder of the right-hand side is a bit of a mess, and contains projections of the non-linear conservative parts as well as the non-conservative parts on $\langle d_{n'}|$;
note that these terms will in general mix different modes $|d_n\rangle$ and $|d_m\rangle$. Rather than writing out explicitly here what the terms look like, we will instead systematically analyze the terms of different order in $c_i$. The general procedure we will follow is to expand all $c_i$'s in the basis $|d_n\rangle$, and then multiply both sides of the equation with ${\rm e}^{i\omega_{n'}t}$ and integrate over time. The crucial assumption now is that the time dependence of $A_n(t)$ is sufficiently slow that it can be held constant during the time integration and moved outside of the
integral, while for the integration over exponential factor, we can use Eq.~(\ref{eqn:freq_comp}). This means we can also
ignore correlations between $A_n(t)$ and $A_{n'}(t)$ on times of the order of $\omega_m^{-1}$ for all
$n,n',m$ while the projected solutions onto modes 1 and 2 will couple $A_1(t)$ and $A_2(t)$ with temporal correlations on time scales much larger than $\omega_m^{-1}$ for 
all $m$. Each time-integration will then
give rise to a condition on the frequency components. These kinds of terms with the condition on the frequencies are entirely
analogous to a magnon-magnon scattering vertex, with the condition on the frequencies corresponding to energy conservation
at the vertex. 
In spin wave Hamiltonian theory, the Holstein-Primakoff transformation is usually applied, and a subsequent expansion of $\sqrt{1-\hat n_i}$, where $\hat n_i$ is the(bosonic) spin number operator at site $i$, leads to magnon interactions
in which magnons scatter off each other. Spin torque oscillators typically have large enough amplitudes that nonlinear processes
are important and have to be included. Therefore, the 
expansions in $|c_i|$ and subsequent expansions of $c_i$ in eigenmode amplitudes $|A_n|$ have to go beyond linear terms, and processes that involve more than two quanta of spin waves have to be considered. This is how linear terms like $dA_{1}/dt\propto A_2(t)$ can arise
as nonlinear processes that include a bath of magnons give some flexibility in satisfying energy conservation in the scattering processes. We shall here expand only up to third order although we will discuss fourth-order contributions that enter because of the 
terms in $a_J\sin\beta$, as well some higher-order contributions. 
For ease of notation, we shall also assume that the demagnetizing tensor is diagonal. This does impact the general conclusions, but it makes the notation a little simpler to follow. Finally, we will assume that all modes $n$ other than modes 1 and 2 are in thermal equilibrium and their
populations can be described by an equilibrium Bose-Einstein distribution function $n_B(\omega_n)$. 
This implies that we are assuming that scattering events
between modes 1 or 2 and other modes $m,n\not=1,2$ are infrequent compared to the equilibration time of modes $m,n$.
\subsection{Linear non-conservative terms}
The linear non-conservative term is
\begin{equation}
-\frac{1}{2}\alpha\left(H_{d,i,y}-iH_{d,i,z}\right)-c_i\left[\alpha\left(H_{\rm ext}+\overline H_{d,i,x}\right)-a_J\cos\beta\right].
\end{equation}
Expanding in eigenmodes, projecting with $\langle d_{n'}|$, multiplying by ${\rm e}^{-i\omega_{n'}t}$ and integrating over time simply leaves 
\begin{equation}
\left(-\alpha\omega_{n'}+a_J\cos\beta\right) A_{n'}(t).\label{eqn:linear_damping}
\end{equation}
As expected, to first order the damping (given by $\alpha$) is offset by the effective pumping (given by $a_J\cos\beta$).
\subsection{Cubic conservative terms}
The cubic conservative terms are
\begin{eqnarray}
&&-\frac{i}{4}\sum_{i'}\left\{D_{i,i';y,y}\left[(c_{i'}+c_{i'}^*)(|c_{i'}|^2+2|c_i|^2+c_i^2)\right]
+D_{i,i';z,z}\left[(c_{i'}-c_{i'}^*)(|c_{i'}|^2+2|c_i|^2-c_i^2\right]\right\}+2ic_i\sum_{i'}D_{i,i';xx}|c_{i'}|^2\nonumber\\
\label{eqn:cubic_cons_1}
\end{eqnarray}
We insert the expansion $c_{i'}=\sum_n A_n(t)d_{i',n}{\rm e}^{-i\omega_n t}$, multiply on the left by $d_{i,n'}^*{\rm e}^{i\omega_{n'}t}$,
sum over $i$ and integrate over $t$ (ignoring the time dependence of $A(t)$). The cubic terms give rise to different possibilities of mode combinations. The terms in $|c_i|^2$ and $|c_{i'}|^2$ give rise to terms like
\begin{equation}
\sum_{n,m,m'}A_{n}A_mA^*_{m'}{\rm e}^{-i (\omega_n+\omega_m-\omega_{m'}-\omega_{n'})t}
\pm\sum_{n,m,m'}A_{n}^*A_mA^*_{m'}{\rm e}^{-i (-\omega_n+\omega_m-\omega_{m'}-\omega_{n'})t},
\label{eqn:cubic_cons}
\end{equation}
which give non-zero contributions if $\omega_n+\omega_m-\omega_{m'}-\omega_{n'}=0$ for the first sum, or
$-\omega_n+\omega_m-\omega_{m'}-\omega_{n'}=0$ for the second one. The different possibilities coupling modes 1 and
2 according to the first sum are depicted in Fig.~\ref{fig:Magnon_scatt}. The special case $n=n'=m=m'=1$ 
gives the non-linear frequency shift for a single-mode theory (not depicted in Fig.~\ref{fig:Magnon_scatt}).
\begin{figure}
\includegraphics*[width=.45\textwidth]{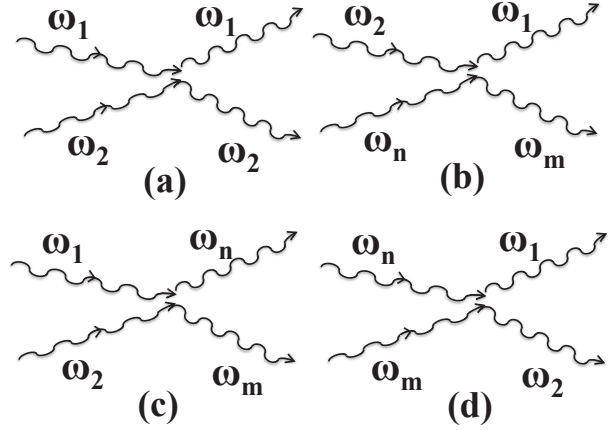}
\caption{Depiction of scattering events coupling modes 1 and 2.}\label{fig:Magnon_scatt}
\end{figure}
For a two-mode theory with $n'=m=1$ and $n=m'=2$ we obtain the off-diagonal non-linear frequency shift of Muduli, Heinonen, and {\AA}kerman\cite{muduli2012prl} [Fig.~\ref{fig:Magnon_scatt} (a)]. Scattering processes such as the one depicted in Fig.~\ref{fig:Magnon_scatt} (b)
will give rise to linear terms of the form $dA_{1}(t)/dt\propto BA_{2}(t)$, with $B$ some constant, provided 
$\omega_1+\omega_m=\omega_2+\omega_n$, or stated differently, provided $\omega_2-\omega_1=\omega_m-\omega_n$, can be satisfied,
and (ii) mode $n$ is
occupied. The latter condition is generally satisfied at finite temperatures due to thermal occupation. The former can in general be satisfied because
firstly, the magnon energies are small compared to room temperature or elevated temperatures in the STOs because of Joule heating. Therefore,
modes in a very large portion of the magnon spectrum are thermally occupied. Secondly, there will then be magnon modes $m$ and $n$ somewhere
in the thermally occupied part of the spectrum such that the energy conservation requirement $\omega_2-\omega_1=\omega_m-\omega_n$
can be satisfied as the spacing between magnon frequencies is not uniform.
The possible contributions to this kind of linear coupling also increase with increasing
order in $c_i$. For example, in fifth order incoming modes $\omega_2$ and $\omega_n$ can scatter into modes 
$\omega_1$, $\omega_m$, $\omega_{m'}$ and $\omega_{m''}$, provided 
$\omega_2+\omega_n=\omega_1+\omega_m+\omega_{m'}+\omega_{m''}$. The contributions to this linear coupling also
increases as a mode crossing is approached so that
$\omega_1\to\omega_2$, in which case $\omega_m\to\omega_n$ and energy conservation is satisfied for any mode $\omega_n$. 
Other possible scattering events coupling modes 1 and 2 are depicted in Fig.~\ref{fig:Magnon_scatt} (c) and (d), but the contributions
from such events are zero because under the assumptions that modes 1 and 2 have the lowest frequencies, these events cannot satisfy
energy conservation at the vertex. Similarly, the second sum in Eq.~(\ref{eqn:cubic_cons}) correspond to one incoming mode and three
outgoing modes at the vertex for which energy can not be conserved at the vertex with $n'=1$ (2) and $n=2$ (1).

The terms in $c_i^2$ in Eq.~(\ref{eqn:cubic_cons_1}) give rise to contributions of the form
\begin{equation}
\sum_{n,m,m'}A_{n}A_mA_{m'}{\rm e}^{-i (\omega_n+\omega_m+\omega_{m'}-\omega_{n'})t}
\pm\sum_{n,m,m'}A_{n}^*A_mA_{m'}{\rm e}^{-i (-\omega_n+\omega_m+\omega_{m'}-\omega_{n'})t},
\label{eqn:cubic_cons_2}
\end{equation}
which, after integrating over time, will give rise to non-zero terms provided $\omega_n+\omega_m+\omega_{m'}-\omega_{n'}=0$, or
$-\omega_n+\omega_m+\omega_{m'}-\omega_{n'}=0$. The former cannot be satisfied for events coupling modes 1 and 2 since there are
three incoming modes, two of which have higher frequencies than modes 1 or 2. The second sum in Eq.~(\ref{eqn:cubic_cons_2}) gives rise to
the same kind of terms as the first sum in Eq.~(\ref{eqn:cubic_cons}).

For an effective two-mode theory with modes 1 and 2 dominant, the diagonal and off-diagonal
nonlinear frequency shifts of mode 1 are given by
\begin{eqnarray}
&&-\frac{i}{4}A_{1}|A_{1}|^2\sum_{i,i'}\left(D_{i,i';yy}+D_{i,i';zz}\right)d_{i,1}^*d_{i',1}|d_{i',1}|^2\nonumber\\
&&-\frac{i}{2}A_{1}|A_{1}|^2\sum_{i,i'}\left(D_{i,i';yy}+D_{i,i';zz}\right)d_{i,1}^*d_{i',1}|d_{i,1}|^2\nonumber\\
&&-\frac{i}{4}A_{1}|A_{1}|^2\sum_{i,i'}\left(D_{i,i';yy}-D_{i,i';zz}\right)d_{i,1}^*d_{i',1}^*d_{i,1}^2\nonumber\\
&&+2iA_{1}|A_{1}|^2\sum_{i,i'}D_{i,i';xx}d_{i,1}^*d_{i',1}|d_{i',1}|^2\nonumber\\
&&-\frac{i}{4}A_{1}|A_{2}|^2\sum_{i,i'}\left(D_{i,i';yy}+D_{i,i';zz}\right)d_{i,1}^*d_{i',1}|d_{i',2}|^2\nonumber\\
&&-\frac{i}{2}A_{1}|A_{2}|^2\sum_{i,i'}\left(D_{i,i';yy}+D_{i,i';zz}\right)d_{i,1}^*d_{i',1}|d_{i,2}|^2\nonumber\\
&&-\frac{i}{4}A_{1}|A_{2}|^2\sum_{i,i'}\left(D_{i,i';yy}-D_{i,i';zz}\right)d_{i,1}^*d_{i',1}^*|d_{i,2}|^2\nonumber\\
&&+2iA_{1}|A_{2}|^2\sum_{i,i'}D_{i,i';xx}|d_{i,1}|^2|d_{i',2}|^2
\label{eqn:nonlinear_f}
\end{eqnarray}
We can write Eq.~(\ref{eqn:nonlinear_f}) symbolically as
\begin{equation}
\frac{dA_{1}(t)}{dt}  = -i\left[\omega_1\eta_{1,1}|A_1|^2+\omega_2|A_2|^2\right]A_1,
\end{equation}
with a similar equation for $dA_2(t)/(dt)$.

The other modes $n$ and $m$ constitute a bath of thermally populated modes. For the case $\omega_{n}+\omega_{2}-\omega_{m}-\omega_{1}=0$
we get "back-scattering" terms of the form
\begin{eqnarray}
\frac{dA_{1}(t)}{dt} & = & -\frac{i}{4}A_{2}\sum_{i,i',m,n}d^*_{i,1}d_{i',n}D_{i,i';yy}A_{n}A^*_{m}d^*_{i',m}d_{i',2}\nonumber\\
                                 &&-2\frac{i}{4}A_{2}\sum_{i,i',m,n}d^*_{i,1}d_{i,n}D_{i,i';yy}A_{n}A^*_{m}d^*_{i,m}d_{i',2}\nonumber\\
                                   &&-\frac{i}{4}A_{2}\sum_{i,i',m,n}d^*_{i,1}d_{i',n}D_{i,i';zz}A_{n}A^*_{m}d^*_{i',m}d_{i',2}\nonumber\\
                                 &&-2\frac{i}{4}A_{2}\sum_{i,i',m,n}d^*_{i,1}d_{i,n}D_{i,i';zz}A_{n}A^*_{m}d^*_{i,m}d_{i',2}\nonumber\\
&=&                                -\frac{i}{4}A_{2}\sum_{i,i',m,n}d^*_{i,1}d_{i',n}D_{i,i';yy}n_B(\omega_n)d^*_{i',m}d_{i',2}\nonumber\\
                                 &&-2\frac{i}{4}A_{2}\sum_{i,i',m,n}d^*_{i,1}d_{i,n}D_{i,i';yy}n_B(\omega_n)d^*_{i,m}d_{i',2}\nonumber\\
                                   &&-\frac{i}{4}A_{2}\sum_{i,i',m,n}d^*_{i,1}d_{i',n}D_{i,i';zz}n_B(\omega_n)d^*_{i',m}d_{i',2}\nonumber\\
                                 &&-2\frac{i}{4}A_{2}\sum_{i,i',m,n}d^*_{i,1}d_{i,n}D_{i,i';zz}n_B(\omega_n)d^*_{i,m}d_{i',2},
\label{eqn:cons_back_scattering}
\end{eqnarray}
where we have used the assumed thermal equilibrium of modes $m$ and $n$ to replace $A_n$ and $A^*_m$ with $n_B(\omega_n)$ and
unity, respectively, by taking a thermal average over magnon states. 
We note that both the nonlinear frequency shift as well as the linear "back-scattering" term are driven by
{\em non-local} nonlinear interactions inherent in the LLG equation.

\subsection{Cubic non-conservative terms}
Just like for the conservative terms, the expansion of $1/\sqrt{1-|c_i|^2}$ in the denominator and $\sqrt{1-|c_{i'}|^2}$ in the field terms
$H_{d,i',y}$ and $H_{d.i',z}$, together with the factors $|c_i|^2$ and $c_i^2$ will give rise to four-magnon vertex 
terms of the same structures as the 
conservative third-order terms and under the same conditions on the magnon frequencies. As these terms arise from the 
non-conservative torques, they will give rise to non-linear damping and pumping with both diagonal terms, as in the
single-mode theory, and off-diagonal terms as in the two-mode description by Mudulu, Heinonen, and Akerman\cite{muduli2012prl}. Finally, there will also in an effective two-mode theory be  a linear contribution
of the form $dA_{1}(t)/dt\propto A_{2}(t)$ if there exist modes $m,n$ such that $\omega_n+\omega_2-\omega_{m}-\omega_{1}=0$.
Again, this term can contribute only when there is a bath of thermally excited modes $n$.

With $\alpha=\alpha_G(1+q_1|c_i|^2)$, the total cubic non-conservative terms are
\begin{eqnarray}
&&\frac{\alpha_G}{4}\sum_{i'}D_{i,i';yy}\left(c_{i'}+c_{i'}^*\right)\left[3c_i^2-2|c_i|^2-|c_{i'}|^2\right]\nonumber\\
&&-\frac{\alpha_G}{4}\sum_{i'}D_{i,i';zz}\left(c_{i'}-c_{i'}^*\right)\left[3c_i^2+2|c_i|^2+|c_{i'}|^2\right]\nonumber\\
&&+2\alpha_Gc_i\sum_{i'}D_{i,i';xx}\left[|c_i|^2-|c_{i'}|^2\right]-a_Jc_i|c_i|^2\cos\beta\nonumber\\
&&+q_1\frac{\alpha_G}{2}\sum_{i'}\left[\left(c_{i'}+c_{i'}^*\right)D_{i,i';yy}+\left(c_{i'}-c_{i'}^*\right)D_{i,i';zz}\right]|c_i|^2\nonumber\\
&&-q_1\alpha_Gc_i|c_i|^2\left(H_{\rm ext}-\overline H_{d,i,x}\right).
\label{eqn:cubic_noncons}
\end{eqnarray}
Again, for two dominant modes $1$ and $2$ we consider resonant scattering corresponding to Fig.~\ref{fig:Magnon_scatt} (a), and get the following contributions to the diagonal and
off-diagonal non-linear damping and pumping:
\begin{eqnarray}
&&\frac{\alpha_G}{2}A_{1}|A_{1}|^2\sum_{i,i'}d_{i,1}^* d_{i',1}|d_{i,1}|^2
D_{i,i';xx}
\nonumber\\
&&  -\frac{\alpha_G}{4}A_{1}|A_{1}|^2\sum_{i,i'} d_{i,1}^*d_{i',1}|d_{i',1}|^2
\left[8D_{i,i';xx}-D_{i,i';yy}-D_{i,i';zz}\right]
\nonumber\\
&& +\frac{\alpha_G}{4}A_{1}|A_{1}|^2\sum_{i,i'} |d_{i,1}|^2d_{i',1}^*d_{i,n}
\left[3D_{i,i';yy}+3D_{i,i';zz}\right]
\nonumber\\
&&-q_1\alpha_GA_{1}|A_{1}|^2\sum_i|d_{i,1}|^2|d_{i,1}|^2\left(H_{\rm ext}-\overline H_{d,i,x}\right)
-a_J\cos\beta A_{1}|A_{1}|^2\sum_i|d_{i,1}|^2|d_{i,1}|^2\nonumber\\
&& +\frac{\alpha_G}{2}A_{1}|A_2|^2\sum_{i,i'} d_{i,1}^*d_{i',1}|d_{i,2}|^2 D_{i,i';xx}\nonumber\\
&& -\frac{\alpha_G}{4}A_{1}|A_2|^2\sum_{i,i'} d_{i,1}^*d_{i',1}|d_{i',2}|^2
\left[8D_{i,i';xx}-D_{i,i';yy}-D_{i,i';zz}\right]\nonumber\\
&& +\frac{\alpha_G}{4} A_{1}|A_2|^2\sum_{i,i'} d_{i,1}^*d_{i',1}^*d_{i,2}^2
\left[3D_{i,i';yy}+3D_{i,i';zz}\right]\nonumber\\
&&-q_1\alpha_GA_{1}|A_2|^2\sum_i|d_{i,1}|^2|d_{i,2}|^2\left(H_{\rm ext}-\overline H_{d,i,x}\right)
-a_J\cos\beta A_{1}|A_2|^2\sum_i|d_{i,1}|^2|d_{i,2}|^2
\label{eqn:cubic_noncons_2}
\end{eqnarray}
By considering scattering with thermally populated modes, as outlined earlier, we can also generate contributions to the "back-scattering" terms analogous
to Eq.~(\ref{eqn:cons_back_scattering}), but we will not write these out here.

In the terms examined so far, linear non-conservative and cubic conservative and non-conservative terms, $a_J\cos\beta$ enters only
as a product. As we indicated earlier, this implies that these equations are invariant under keeping $a_J\cos\theta$ constant, which can
only explain an increase with threshold current with decreasing $\cos\beta$. It cannot explain any other behavior that changes with
angle even if $a_J\cos\beta$ is held constant. Therefore, such angular dependence can only come from the terms in $a_J\sin\beta$ in
Eq.~(\ref{eqn:dc_dt_3}).

We can combine the linear [Eq.~(\ref{eqn:linear_damping})] and cubic nonconservative damping and pumping terms [Eq.~(\ref{eqn:cubic_noncons_2})], and also the 
conservative and non-conservative "back-scattering" terms in the following equation
\begin{equation}
\frac{dA_{1}(t)}{dt} = -\Gamma_G\left[1+P_{1,1}\omega_1|A_1|^2
+P_{1,2}\omega_2|A_2|^2\right]A_1
+\sigma_0I\cos\beta\left[1-Q_{1,1}\omega_1|A_1|^2-Q_{1,2}\omega_2|A_2|^2\right]A_1+R_{1,2}(T)A_2,
\end{equation}
with a similar equation for the nonlinear damping and pumping, and back-scattering, contributions to $dA_2(t)/(dt)$. We have
noted the temperature dependence of the linear term that arises from scattering of thermally populated modes.
In contrast with the nonlinear frequency shift, the non-linear damping and pumping is driven both by local (lines 5 and 8 in Eq.~(\ref{eqn:cubic_noncons_2}) as well as non-local terms; of course, the nonlinear damping originates in the magnetostatic interactions, just like
the nonlinear frequency shift.  

The thermal populations of modes $n$ and $m$ contributing to 
back-scattering terms arising from cubic terms have as a consequence that
the backscattering terms will have a direct temperature dependence. As the backscattering terms are responsible for bifurcation and
mode crossing\cite{beri2008prl,vanderSande} this implies that the manifold of periodic orbits and fixed points, both stable and unstable ones at
saddle points, will shift as the temperature is varied. This is in contrast to the temperature effects that occur when the system is coupled
to a thermal bath that gives rise to a stochastic field. Thermal fluctuations will primarily induce mode-hopping over saddle points and thermal excursions around periodic orbits and stable fixed points, but have a small effect on the manifold itself.


\subsection{$\sin\beta\not= 0$}
For $\sin\beta\not=0$, that is, when the fixed layer direction is not aligned with the equilibrium magnetization direction of the free layer, there
arise new terms of different symmetry than what is otherwise the case: the terms in $\sin\beta$ are all in even powers of $|c_i|$ or
$c_{i}$. This is in contrast with the other terms (both conservative and non-conservative ones), that all have odd powers in $c_i$ or
$|c_{i'}|$. Therefore, these terms in $\sin\beta$ cannot be canceled by the other conservative or non-conservative terms. As a consequence, new mode-mode scattering channels open up when the applied external field is
rotated away from the direction of the fixed layer magnetization. Again, considering the two dominant modes 1 and 2, the lowest-order contributions from the terms in $c_i^2$ coupling modes 1 and 2 occur if
$\omega_{2}+\omega_{m}=\omega_{1}$, or $\omega_1+\omega_m=\omega_2$, or $\omega_1+\omega_2=\omega_m$ 
is satisfied for some $m$. The first two of these are not allowed as $\omega_m>\omega_2,\omega_1$,
but the last represents the scattering of a magnon pair of modes 1 and 2 into mode $m$. 
Similarly the lowest-order term in $|c_i|^2$
only gives a non-zero contributions if $\omega_{m}=\omega_{2}+\omega_{1}$, which represents a decay of a thermally populated mode
$m$ into modes 1 and 2. 
The available phase-space that satisfies 
the requirement $\omega_{m}=\omega_{2}+\omega_{1}$ for the possible 
three-magnon processes is at the most satisfied at special discrete values of an external control parameter,
such as external field magnitude or direction, and will therefore be ignored here. Higher-order terms have larger
available phase-space. In forth order, we have the terms
\begin{equation}
\frac{1}{8}a_J\sin[\beta]\left[|c_i|^4+c_i^2|c_i|^2\right].
\end{equation}
We consider only contributions to $dA_1/(dt)$ which in diagrams of the type in Fig.~\ref{fig:Magnon_scatt} have at least
one outgoing mode 1 magnon. Also, we only include scattering events that are compatible with $\omega_m >\omega_1,\omega_2$ for all
$m\not=1,2$. Finally, we exclude scattering events with only one magnon in the thermal bath as the requirement on energy conservation
at the vertex can in general not be satisfied for such events. In all, we get the following contributions to $dA_1/(dt)$:
\begin{eqnarray}
\frac{1}{8}a_j\sin[\beta]&&\left\{\sum_{m,n,i}A_1A_2n_B(\omega_n)\delta(\omega_n+\omega_2-\omega_m)|d_{i,1}|^2d_{i,2}d_{i,m}^*d_{i,n}\right.
\nonumber\\
&&\sum_{m,n,i}A_2^2n_B(\omega_n)\delta(\omega_n+2\omega_2-\omega_m-\omega_1)d_{i,1}^*d_{i,2}^2d_{i,m}^*d_{i,n}\nonumber\\
&&\sum_{m,n,i}|A_2|^2n_B(\omega_n)\delta(\omega_n-\omega_m-\omega_1)d_{i,1}^*|d_{i,2}|^2d_{i,m}^*d_{i,n}\nonumber\\
&&\sum_{m,n,i}A_1^*A_2n_B(\omega_n)\delta(\omega_n+\omega_2-\omega_m-2\omega_1){d_{i,1}^*}^2d_{i,2}d_{i,m}^*d_{i,n}\nonumber\\
&&\sum_{m,n,i}A_1^2n_B(\omega_n)\delta(\omega_n+\omega_1-\omega_m)|d_{i,1}|^2d_{i,1}d_{i,m}^*d_{i,n}\nonumber\\
&&\left.\sum_{m,n,i}|A_1|^2n_B(\omega_n)\delta(\omega_n-\omega_1-\omega_m)|d_{i,1}|^2d_{i,1}^*d_{i,m}^*d_{i,n}\right\}.
\end{eqnarray}
These terms are more complicated than the ones we have considered previously in that they will not only couple the time evolution
of $A_1$ and $A_2$ with linear terms or terms of the form of the non-linear damping or pumping, and their inclusion in an 
effective theory would require a much larger parameter set. We will not here further discuss such an effective theory as
we are assuming that an expansion to third orded is sufficient, and therefore fourth-order terms multiplying $\sin\beta$ can 
certainly be ignored for small $\beta$. The main reason
for discussing these terms is to point out that they alone can cause dependence on angle $\beta$ other than $a_J\cos\beta$ and are,
for example, responsible for the observed increase in mode-hopping as $\cos\beta$ is decreased.


\subsection{General equation for mode coupling}
We can now collect all the terms up to and including third-order 
and write
the coupled equations for modes 1 and 2 as
\begin{eqnarray}
\frac{dA_1(t)}{dt} 
&=& -i\left[\omega_1\eta_{1,1}|A_1|^2+\omega_2\eta_{1,2}|A_2|^2\right]A_1\nonumber\\
&&-\Gamma_G\left[1+P_{1,1}\omega_1|A_1|^2
+P_{1,2}\omega_2|A_2|^2\right]A_1\nonumber\\
&&+\sigma_0I\cos\beta\left[1-Q_{1,1}\omega_1|A_1|^2-Q_{1,2}\omega_2|A_2|^2\right]A_1+R_{1,2}(T)A_2
\label{eqn:LLG_reduce_2_1}\\
\frac{dA_2(t)}{dt} 
&=&-i\left[\omega_1\eta_{2,1}|A_1|^2+\omega_2\eta_{2,2}|A_2|^2\right]A_2\nonumber\\
&&-\Gamma_G\left[1+P_{2,1}\omega_1|A_1|^2+P_{2,2}\omega_2|A_2|^2\right]A_2\nonumber\\
&&+\sigma_0 I \cos\beta\left[1-Q_{2,1}\omega_1|A_1|^2-Q_{2,2}\omega_2|A_2|^2
\right]A_2+R_{2,1}(T)A_1
\label{eqn:LLG_reduce_2_2}
\end{eqnarray}
Equations~(\ref{eqn:LLG_reduce_2_1}) and (\ref{eqn:LLG_reduce_2_2}) are the main result of the present work. They are the equations given by Muduli, Heinonen, and {\AA}kerman\cite{muduli2012prl}.

\section{Summary and Conclusions}
We have here presented detailed derivations of the equations of motions for coupled modes in STOs. In particular, we have shown
that the equations governing a system with two dominant modes can be reduced to a set of coupled equations first given by
Muduli, Heinonen, and {\AA}kerman\cite{muduli2012prl}, and are a generalization of the equations governing a single-mode STO, as given by 
Slavin and Tiberkevich\cite{slavin2008ieeem,slavin2009ieeem}.  We have given explicit expressions for the linear terms and for the cubic 
terms responsible for nonlinear damping and pumping, as well as for the nonlinear frequency shift, 
for the geometries considered here. The linear
"back-scattering" term arise from scattering that is possible when
there is a bath of modes available, and we have given explicit examples of such terms.
In practice, these terms are difficult to calulate and the corresponding coefficients, $R_{1,2}$,
and $R_{2,1}$, can probably be treated as parameters in modeling\cite{heinonen2013ieee}. We have also
concluded that these back-scattering terms have a direct temperature dependence as they involve thermal populations
of modes. This implies that manifold of of orbit and fixed point will shift as a function of temperature, in addition to temperature
effects, such as mode hopping, that may be the consequence of a stochastic field that arises from coupling to a thermal bath. 
The equations for the coupled modes include additional terms beyond third order that arise when the free layer equilibrium 
magnetization is
not aligned with the fixed layer magnetization. These terms generate additional coupling between modes that provide
a physical mechanism for observed 
increased mode-hopping as the external field is moved away from the direction of the fixed layer magnetization.
The intrinsic nonlinear and
non-local interactions in the LLG equation that governs the magnetization dynamics give rise to couplings between modes that in turn
can generate a wealth of interesting complicated phenomena, such as mode-hopping or mode coexistence. These couplings and
ensuing phenomena are of fundamental interest
but also of significant importance for technological applications of STOs. 

Yan Zhou acknowledges the support by the Seed Funding Program for Basic Research from the University of Hong Kong, and University Grants Committee of Hong Kong (Contract No. AoE/P-04/08). Argonne National Laboratory is operated under Contract No. DE-AC02-06CH11357 by UChicago Argonne, LLC. Comments by E. Iacocca are greatly appreciated.

\appendix{\bf APPENDIX}

Here, we present the details of the derivation of Eq.~(\ref{eqn:single_mode_3}) from Eq.~(\ref{eqn:single_mode_2}). The linearized version of Eq.~(\ref{eqn:single_mode_2})
is
\begin{equation}
\frac{dc}{dt}=-i\left( bc+nc^*\right),\label{eqn:App_A_linear}
\end{equation}
where $b=H_{\rm ext}-X+\frac{1}{2}\left(Y+Z\right)$, and $n=\frac{1}{2}\left(Y-Z\right)$  
with eigenvalue $\omega$ given by $\omega^2=b^2-n^2$. Equation~(\ref{eqn:App_A_linear}) is diagonalized by introducing a complex variable
$a$ such that $c=ua-va^*$ where $u=\sqrt{(b+\omega)/(2b)}$ and $v=\sqrt{(b-\omega)/(2b)}$. Thus $u$ and $v$ are real with
$u^2+v^2=1$ and $u^2-v^2=\omega/b$. The variable $a$ is given in terms of $c$ by $a=(uc+vc^*)/(u^2-v^2)$. We multiply
Eq.(\ref{eqn:single_mode_2}) by $u/(u^2-v^2)$ and add from the result the complex conjugate of Eq.~(\ref{eqn:single_mode_2}) 
multiplied by $v/(u^2-v^2)$. The linear terms then become
\begin{equation}
\frac{da}{dt}=-i\omega a\left(1-i\alpha\right)
\end{equation}
To work out the rest, we need
\begin{eqnarray}
|c|^2 & = & |a|^2-\frac{n}{2b}\left( a^2+{a^*}^2\right)\label{eqn:App_A_1}\\
c^2   & = & \frac{b+\omega}{2b}{a}^2+\frac{b-\omega}{2b}{a}^{*2}-\frac{n}{b}|a|^2\label{eqn:App_A_2}\\
{c^*}^2 & = & \frac{b+\omega}{2b}{{a}*}^2+\frac{b-\omega}{2b}a^2-\frac{n}{b}|a|^2\label{eqn:App_A_3}\\
\frac{uc-vc^*}{u^2-v^2} & = & \frac{b}{\omega}a-\frac{n}{\omega}a^*.\label{eqn:App_A_4}
\end{eqnarray}
The most tedious part is the second line of Eq.~(\ref{eqn:single_mode_2}), which is
\begin{equation}
-\frac{i}{2}\left(3H_{\rm ext}+X\right)c|c|^2+\frac{i}{4}\left(Y-Z\right)c^3+\frac{i}{4}\left(Y+Z\right)c|c|^2
=-iAc|c|^2+\frac{i}{2}nc^3,\end{equation}
where $A=\frac{1}{2}\left(3H_{\rm ext}+X\right)-\frac{1}{2}(Y+Z)$. We therefore need to evaluate
\begin{equation}
-iA\frac{u}{u^2-v^2}c|c|^2+iA\frac{v}{u^2-v^2}c^*|c|^2+\frac{1}{2}\frac{inu}{u^2-v^2}c^3-\frac{1}{2}\frac{inv}{u^2-v^2}{c^*}^3.
\end{equation}
Using Eqs.~(\ref{eqn:App_A_1}) and (\ref{eqn:App_A_4}), we obtain
\begin{equation}
\frac{uc-vc^*}{u^2-v^2}|c|^2=\frac{2b^2+n^2}{2b\omega}a|a|^2+\frac{3n}{2\omega}{a}^*|{a}|^2-\frac{n}{2\omega}a^3
+\frac{n^2}{2b\omega}{a}^{*3}.
\end{equation}
From Eqs.~(\ref{eqn:App_A_2}-\ref{eqn:App_A_4}) we obtain
\begin{equation}
\frac{nuc}{u^2-v^2}c^2-\frac{nvc^*}{u^2-v^2}{c^*}^2 = 
\frac{3}{2}\frac{n^3}{b\omega}{a^*}|a|^2-\frac{3}{2}\frac{bn^2}{b\omega}a|a|^2+
\frac{1}{2}\frac{n(b^2+\omega^2)}{b\omega}a^3-\frac{1}{2}\frac{n^2b}{b\omega}{a}^{*3}.
\end{equation}
and also
\begin{equation}
\frac{u}{u^2-v^2}c^3+\frac{v}{u^2-v^2}{c^*}^3  =  a^3+\frac{n}{2b}{a}^{*3}-\frac{3n}{2b}a|a|^2.
\label{eqn:uc3+vc*3}
\end{equation}
Putting it all together, we get
\begin{eqnarray}
\frac{da}{dt} & = & -i\omega a\left[1-i\alpha\right]+\frac{3i\omega}{2}a
\left[|a|^2-\frac{n}{2b}\left( a^2+{a^*}^2\right)\right][1-i\alpha]\nonumber\\
&&-iA\left[
\frac{2b^2+n^2}{2b\omega}a|a|^2-\frac{3n}{2\omega}a^*|a|^2-\frac{n}{2\omega}a^3
+\frac{n^2}{2b\omega}{a}^{*3}
\right]\nonumber\\
&&+\frac{i n}{2}\left[
\frac{3n^2}{2b\omega}a^*|a|^2-\frac{3n}{2\omega}a|a|^2+\frac{b^2
+\omega^2}{2b\omega}a^3-\frac{n}{2b\omega}{a}^{*3}
\right]
\nonumber\\
&&-\frac{\alpha}{2}\left[ H_{\rm ext}+3X-3m\right]\left[
a|a|^2-\frac{n}{2b}\left( a^3+a^*|a|^2\right)\right]\nonumber\\
&&+\frac{3\alpha n}{2}\left[ a^3+\frac{n}{2b}{a}^{*3}-\frac{3n}{2b}a|a|^2.\right]
\nonumber\\
&&+\frac{a_J}{2}\frac{b}{\omega}\sin\beta\sqrt{\frac{b+n}{b}}\left[1+\frac{3n-2b}{2b}|a|^2+\frac{5n-6b}{4b}a^2
-\frac{n}{4b}{a}^{*2}\right]
\nonumber\\
&&+a_J\cos\beta a\left[
1-|a|^2+\frac{n}{2b}\left( a^2-{a^*}^2\right)\right].
\end{eqnarray}

\bibliography{modecoupling_v2}
\bibliographystyle{apsrev}
\end{document}